\documentclass[12pt]{article}
\usepackage[]{epsfig}
\usepackage[centertags]{amsmath}
\usepackage{amsfonts}
\usepackage{amssymb}
\usepackage[english]{babel}

\newcommand{\diag}{\text{diag}}%
\newcommand{\Id}{\text{Id}}%

\begin{document}
%%%%%%%%%%%%%%%%%%%%%%%%%%%%%%%%%%%%%%%%%%%%%%%%%%%%%%%%%%%%%%%%%%%%%
\title{\bf Non-Abelian Chern-Simons Vortices}
%%%%%%%%%%%%%%%%%%%%%%%%%%%%%%%%%%%%%%%%%%%%%%%%%%%%%%%%%%%%%%%%%%%%%
\author{G.S.~Lozano$^a$\thanks{Associated with CONICET}\,,
D.~Marqu\'es$^{b,c \, *}$ E.F.~Moreno$^{c,d \, *}$ and
F.A.~Schaposnik$^{b,c}$\thanks{Associated with CICBA}
\\
{\normalsize\it $^a$Departamento de F\'\i sica, FCEyN, Universidad
de Buenos Aires}\\ {\normalsize\it Pab. I, Ciudad Universitaria,
1428, Buenos Aires, Argentina}
\\
{\normalsize\it $^b$Departamento de F\'\i
sica, Universidad Nacional de La Plata}\\ {\normalsize\it C.C. 67,
1900 La Plata, Argentina}
\\
{\normalsize\it $^c$CEFIMAS-SCA, Av. Santa Fe 1145}\\
{\normalsize\it C1059ABF, Buenos Aires, Argentina} \\
{\normalsize\it $^d$Department of Physics, West Virginia
University}\\ {\normalsize\it Morgantown, West Virginia
26506-6315, USA} }

\maketitle
%===================================================================
%===================================================================
\begin{abstract}
We consider the bosonic sector of a ${\cal N} = 2$ supersymmetric
Chern-Simons-Higgs theory in 2 + 1 dimensions. The gauge group is
$U(1)\times SU(N)$ and has $N_f$ flavors of fundamental matter
fields. The model supports non-Abelian (axially symmetric) vortices
when $N_f \geq N$, which have internal (orientational) moduli. When
$N_f > N$, the solutions acquire  additional collective coordinates
parameterizing their transverse size. We solve the BPS equations
numerically and obtain local ($N_f = N$) and semi-local ($N_f > N$)
string solutions.
\end{abstract}

\maketitle

%%%%%%%%%%%%%%%%%%%%%%%%%%%%%%%%%%%%%%%%%%%%%%%%%%%%%%%%%%%%%%%%%%%
\section{Introduction}

Chern-Simons (CS) theories are relevant in a quantum field theory
context since they provide an alternative gauge-invariant procedure
of mass generation \cite{DJT}. Moreover, the high-temperature limit
of quantum field theories in $d=4$ dimensions are effectively three
dimensional and CS terms are precisely induced by fermions in $d=3$
dimensions through the parity anomaly \cite{Redlich}. CS actions
play also a role in the analysis of interesting condensed matter
phenomena \cite{Poly}-\cite{mur},  the computation of topological
invariants of 3-manifolds \cite{witt} and they are connected with
$d=2$ conformal field theories \cite{Elitzur}.

CS-Higgs models differ drastically from theories in which solely a
Maxwell or Yang-Mills term governs the dynamics of the gauge fields.
In particular, at the classical level, axially symmetric (vortex or
string) solutions  to the equations of motion necessarily carry
electric charge \cite{Paul}-\cite{Cugliandolo} which, in the
non-Abelian case, is quantized (for a complete review on CS theories
and planar physics see \cite{Dunne}).

Bogomolny equations for  the non-Abelian CS-Higgs system were first
obtained in \cite{Cugliandolo}, where it was shown that, as in the
Abelian case \cite{HKP}-\cite{JW},  a sixth order potential has to be
considered. Explicit vortex solutions were exhibited in
\cite{Cugliandolo}, with the flux directed in the Cartan subalgebra of
the non-Abelian group. It is the purpose of the present work  to find
genuine non-Abelian vortex configurations by proposing  a more general
ansatz as the one already considered for Yang-Mills-matter theories
\cite{Auzzi}-\cite{GSY} with gauge group $U(1)\times SU(N)$ and $N_f \geq
N$ flavors of fundamental matter multiplets. The resulting vortex
configurations can be characterized by non-Abelian collective coordinates
related to orientational degrees of freedom and, when $N_f > N$, to
infinitesimal variations of the transverse size.

\section{Model and Notation}

We consider the truncated bosonic sector of the  ${\cal N} = 2$
SUSY $U(1) \times SU(N)$ Chern-Simons-Higgs action in $2+1$
dimensions (the complete SUSY Lagrangian can be found in
\cite{GN})
\begin{align}
{\cal S} = \int d^3 x \ &\left\{ \frac{\kappa_1}{2}
\epsilon^{\mu\nu\rho} F_{\mu\nu}^0 A_\rho^0 + \frac{\kappa_2}{2}
\epsilon^{\mu\nu\rho} \left( F_{\mu\nu}^I A_\rho^I - \frac{1}{3}
f^{IJK} A_\mu^I A_\nu^J A_\rho^K \right) \right.+ \nonumber\\
&\left. \left( D_\mu \phi^f
\right)^\dag \left( D^\mu \phi^f \right) - V\left[\phi,
\phi^\dag\right] \right\}, \label{uno}
\end{align}
where $\epsilon^{012} = 1$, $g^{00} = 1$, $f^{IJK}$ are the structure
constants of the non-abelian group, and the quantization condition
implies $\kappa_2 = m / 8 \pi$. Our theory has a sixth-order potential
which, forced by ${\cal N}=2$ supersymmetry, allows Bogomolny completion
of the energy functional with all coupling constants at  the Bogomolny
point (see ref.\cite{FAS} and references therein for details on this
point)
\begin{align}
V\left[\phi, \phi^\dag\right] &= \frac{1}{16 \kappa_1^2 N^2} \phi^\dag_f
\phi^f \left(\phi^\dag_g\phi^g - N\xi\right)^2 +
\frac{1}{4\kappa_2^2}\phi^\dag_f\tau^I\tau^J\phi^f
\left(\phi^\dag_g\tau^I\phi^g\right)
\left(\phi^\dag_h\tau^J\phi^h\right)\nonumber\\
&- \frac{1}{4\kappa_1\kappa_2
N}\left(\phi^\dag_f\tau^I\phi^f\right)^2\left(\phi^\dag_g\phi^g -
N\xi\right).
\end{align}
Here, $\mu, \nu, \rho = 0, 1, 2$ are Lorentz indices, $I, J, K = 1,
\dots, N^2 - 1$ are the $SU(N)$ ``color'' group indices and $\tau_I$ are
the anti-hermitian generators of $SU(N)$. The complex scalar multiplets
$\phi^f_i$, besides the color index $i, j, k = 1, \dots, N$, possess
additional flavor index $f, g, h = 1, \dots, N_f$ with $N_f \geq N$, thus
can be written as  $N \times N_f$ matrices. The covariant derivatives and
field strengths are defined as
\begin{align}
&D_\mu\phi^f_i = \partial_\mu \phi^f_i + (A_\mu^{SU(N)})^j_i \, \phi^f_j
+ (A_\mu^{U(1)})^j_i \, \phi^f_j \; ,\nonumber \\
&A_\mu^{SU(N)} = A_\mu^I \ \tau_I\ ,&
& \hspace{-4cm} A_\mu^{U(1)} = A_\mu^0 \ \tau_0
\nonumber \\
&F_{\mu\nu}^0 = \partial_{[\mu} A_{\nu ]}^0\ , & & \hspace{-4cm}
F_{\mu\nu}^I = \partial_{[\mu} A_{\nu ]}^I + f^{IJK} A_{\mu}^J A_{\nu}^K
\; .
\end{align}
Up to gauge transformations, minima of the potential are given by
\begin{align}
& \text{\sl symmetric  phase}  & &\phi^f = 0\\
& \text{\sl asymmetric  phase}  & &\phi^f\phi^\dag_f = \xi \ \diag\{1,
\ldots, 1\}\ .
\label{Nontrivialvacuum}
\end{align}
In what follows we set, without loss of generality, $\xi = 1$. The energy
density is
\begin{equation}
{\cal H} = \left( D^0 \phi^f \right)^\dag \left( D^0 \phi^f \right)
+ \left( D^i \phi^f \right)^\dag \left( D^i \phi^f \right) + V
\left[\phi, \phi^\dag\right]\
\end{equation}
and the Euler-Lagrange equations of motion of the theory are
\begin{align}
\kappa_1 \, \epsilon_\mu^{\,\alpha\beta}\
F^0_{\alpha\beta}&=J_\mu^0  \equiv  \phi^\dag_f \tau^0 D_\mu
\phi^f - \left(D_\mu
\phi^f\right)^\dag \tau^0 \phi^f \nonumber\\
\kappa_2 \, \epsilon_\mu^{\, \alpha\beta}\
F^I_{\alpha\beta}&=J_\mu^I  \equiv \phi^\dag_f \tau^I D_\mu
\phi^f - \left(D_\mu \phi^f\right)^\dag
\tau^I \phi^f  \label{Gauss} \\
D_\mu D^\mu \phi^f&=\frac{\partial V}{\partial \phi_f^\dag}\ .
\end{align}
Defining $D_\epsilon \equiv D_1 + i \epsilon D_2$ with $\epsilon \equiv
\pm$, and using Gauss' law, we can write the energy as a sum of squares
\begin{align}
H &= \int d^2 x \;  \left\{  \left[ D_0 \phi^f - i\epsilon \left(
\frac{1}{4\kappa_1 N}\left(\phi^\dag_g\phi^g - N \right) \phi^f -
\frac{1}{2\kappa_2} \phi^\dag_g\tau^I\phi^g\tau^I\phi^f
\right)\right]^\dag \right. \nonumber\\
& \times \left[ D_0 \phi^f - i\epsilon \left( \frac{1}{4\kappa_1
N}\left[\phi^\dag_g\phi^g - N \right] \phi^f -\frac{1}{2\kappa_2}
\phi^\dag_g\tau^I\phi^g\tau^I\phi^f
\right)\right]  \nonumber \\
& + \left. \vphantom{\frac{1}{4\kappa_1}}
\left(D_{-\epsilon}\phi^f\right)^\dag \left(D_{-\epsilon} \phi^f\right) +
\epsilon\sqrt{2N} F^0_{12} \right\}
\end{align}
leading to the Bogomolny equations.

\section{Non-abelian local strings}

In this section we investigate the so-called ``local $Z_N$ string-type
solutions'' as discussed for the Yang-Mills case in
\cite{Auzzi}-\cite{GSY} (see also \cite{aldro}).

We set $N_f = N$ so the matter fields can be arranged as a square matrix
$\Phi$. The Lagrangian (\ref{uno}) is then invariant under
$SU(N)_{color}\times SU(N)_{flavor}$ rotations,
\begin{equation}
\Phi \to U \Phi V \; ,
  \;\;\;  \;\;\;  A_\mu \to U A_\mu U^{-1} -( \partial_\mu U )U^{-1}
\end{equation}
with $\,U \in U(N)_{local}\,$ and $V\in SU(N)_{global}$.

We start from the trivial vacuum in the asymmetric phase
\begin{equation}
A^{vac}_\mu = 0\ , \ \ \ \ \ \Phi^{vac} = \diag\{1, \dots, 1\}\ .
\label{vacios}
\end{equation}
After a $U(1)\times SU(N)$ gauge transformation we obtain a singular
vortex configuration which, for the scalar field, takes the form
\begin{equation}
\Phi^{vac} \longrightarrow \Phi = \exp\left(\alpha \tau^0 + \beta
\tau^{N^2-1}\right) \Phi^{vac} = \, \diag\{1, \dots, 1, e^{-i\varphi n
\epsilon}\}
\label{15}
\end{equation}
with
\begin{align}
&\tau^0 = \frac{i}{\sqrt{2N}} \; \diag\{1,\dots,1\}\;, &
&\tau^{N^2-1}= \frac{i}{\sqrt{2N(N-1)}}
\; \diag\{1,\dots,1,1-N\} \label{matrices} \\
&\alpha = -\sqrt{2/N}\; \epsilon n \varphi\;, & &\beta =
\sqrt{2(N-1)/{N}} \; \epsilon n \varphi \nonumber
\end{align}
Concerning the gauge fields we have
\begin{eqnarray}
&& A_i^{vac\, 0} \to A_i^0 =
-\sqrt{\frac{2}{N}}\varepsilon_{ij}\frac{x_j}{r^2}n\epsilon \; ,
\;\;\;\;\;\; \;\;\;\;\;\; A_i^{vac\,N^2\!\!-1 } \to A^{N^2\!\!-1}_i
= \sqrt{\frac{2(N-1)}{N}}\varepsilon_{ij}
\frac{x_j}{r^2}n\epsilon \ , \nonumber\\
&& A_i^{vac\, I}  \to  A_i^{I} = 0  \; , \;\;\;\;\;\;
\;\;\;\;\;\;\;\;\;\;\;\; \;\;\;\;\;\;\;\;\;\;\;\;A_0^{vac\, I} \to  A_0^{I} = 0 \; ,
 \;\;\; I=1,2, \ldots,N^2-2\label{20}
\end{eqnarray}
%
%
%\begin{align}
%A_i^{vac\, 0} &\to A_i^0 =
%-\sqrt{\frac{2}{N}}\varepsilon_{ij}\frac{x_j}{r^2}n\epsilon
%\nonumber\\
%%
%A_i^{vac\,N^2\!\!-1 } &\to A^{N^2\!\!-1}_i =
%\sqrt{\frac{2(N-1)}{N}}\varepsilon_{ij}
%\frac{x_j}{r^2}n\epsilon \ , \nonumber\\
%%
%A_i^{vac\, I} &\to  A_i^{I} = 0  \; , \;\;\; I=1,2, \ldots,N^2-2
%\nonumber\\
%%
%A_0^{vac\, I} &\to  A_0^{I} = 0 \label{20}
%\end{align}
%%
In such configuration, the $Z_N$ center of the gauge group $SU(N)$ has
been combined with $U(1)$ elements to get a topologically stable string
solution possessing both windings, in $SU(N)$ and in $U(1)$ (since $\pi_1
\left({SU(N)\times U(1)}/{Z_N}\right) \neq 0$, the topology is
nontrivial). These kind of topological objects are also called ``$Z_N$
strings". In the present case the configurations (\ref{15}),(\ref{20})
represent a $(0,\ldots,0,n)$ singular string. A general $(n_1,\ldots,
n_N)$ vortex configuration can be obtained following the same method.
This suggests the following ansatz for the regular vortex configuration
\begin{align}
&\Phi = \diag\{\phi(r),\dots,\phi(r),e^{-i\varphi
n\epsilon}\phi_N(r)\} \;, & &\label{fi}\\
&A_0^{N^2-1} = \sqrt{\frac{N-1}{2N}}f_0^{N^2-1}(r) \; ,
&%
&  \hspace{-1cm}A^{N^2-1}_i = \sqrt{\frac{2(N-1)}{N}}\varepsilon_{ij}
\frac{x_j}{r^2}(n\epsilon + f^{N^2-1}(r))\nonumber\\
&A^0_0 = \frac{1}{\sqrt{2N}}f_0(r)\; ,
& %
& \hspace{-1cm} A_i^0 = -
\sqrt{\frac{2}{N}}\varepsilon_{ij}\frac{x_j}{r^2}(n\epsilon + f(r)) \ .
\label{superansatz}
\end{align}
It should be noted that, contrary to the Yang-Mills-Higgs case
\cite{GSY}, the $A_0$ fields are nontrivial. The boundary conditions for
the fields are
\begin{align}
&\phi_N (0) = 0, \ \ \ \ f(0) = f^{N^2-1}(0) = -\epsilon\ n\, ,
\nonumber\\
&\phi(\infty) = \phi_N(\infty) = 1, \ \ \ \ f_0(\infty) =
f_0^{N^2-1}(\infty) = f(\infty) = f^{N^2-1}(\infty) = 0\ ,
\label{boundcond}
\end{align}
where we required the solution to be singled-valued at the origin, and to
be a pure gauge at infinity. With this ansatz both the flux and the
energy of these solutions are quantized
\begin{align}
\Phi \equiv \int d^2x F^0_{12} = \frac{4\pi}{\sqrt{2N}} \,\epsilon
n \; , \;\;\;\;\;
E= 2\pi\, n . \label{flux-energy}
\end{align}
Note that with our conventions $n$ is always a positive integer,
and $\epsilon$ determines whether the flux is positive ($\epsilon
= +$), or negative ($\epsilon = -$).

Ansatz (\ref{fi}) corresponds, for the $n=1$ case, to what is
called an ``elementary string''.  Composite strings can be
constructed by introducing windings in several diagonal elements
in the scalar field and  can be seen as the superposition of
elementary ones.

Substituting ansatz (\ref{fi})-(\ref{superansatz}) into the
Bogomolny and Gauss equations, we obtain the following system of
non-linear first-order differential equations
\begin{align}
&r\partial_r \phi = -\frac{\epsilon}{N}\left(f - f^{N^2-1}\right)\phi \; ,
\;\;\;\; r\partial_r \phi_N = -\frac{\epsilon}{N} \left(f + (N - 1)
f^{N^2-1}\right) \phi_N \label{non1}\\
&\frac{1}{r}\partial_r f = - \frac{1}{4 N \kappa_1} \left[f_0 \left((N -
1) \phi^2 + \phi_N^2\right) + f_0^{N^2-1} (N - 1)\left(\phi^2 -
\phi_N^2\right)\right] \label{non3}\\
&\frac{1}{r}\partial_r f^{N^2-1} = \frac{1}{4 N \kappa_2} \left[f_0
\left(\phi^2 - \phi^2_N\right) + f_0^{N^2-1} \left(\phi^2 + (N - 1)
\phi^2_N\right)\right]\label{non4}\\
&f_0 = \frac{\epsilon}{2\kappa_1}\left((N-1)\phi^2 + \phi_N^2 -
N\right) \; ,
\;\;\;\;\; f_0^{N^2-1} = \frac{\epsilon}{2\kappa_2} \left(\phi^2 -
\phi_N^2\right). \label{non6}
\end{align}
Note that equations (\ref{non6}) can be used to eliminate $f_0$ and
$f_0^{N^2-1}$ in (\ref{non3}) and (\ref{non4}). Equations
(\ref{non1})-(\ref{non4}) then correspond to the Bogomolny equations
written in the standard form. In terms of the profile functions, the
energy takes the form
\begin{align}
E &= 2\pi \int r\ dr \left\{\frac{1}{8
N^2\kappa_1^2}\left((N-1)\phi^2 + \phi_N^2 -
N\right)^2 \right. \left((N-1)\phi^2
+ \phi_N^2\right) \nonumber \\
&+ \frac{(N-1)}{8 N^2\kappa_2^2}\left(\phi^2 -
\phi_N^2\right)^2\left(\phi^2 + (N-1) \phi_N^2\right) +
\frac{(N-1)}{4N^2\kappa_1\kappa_2}\left(\phi^2 -
\phi_N^2\right)^2 \times \nonumber \\
& \left((N-1)\phi^2 + \phi_N^2 -
N\right)+  (N-1) (\partial_r\phi)^2 + (\partial_r\phi_N)^2 \nonumber \\
& \left. + \frac{(N-1)}{r^2N^2}\phi^2\left(f-f^{N^2-1}\right)^2 +
\frac{1}{N^2r^2}\phi_N^2\left(f + (N-1)f^{N^2-1}\right)^2 \right\}
\label{energy2}%
\end{align}
The magnetic field (which is a pseudoscalar in 2+1 dimensions) and the
electric field (with component only in the radial direction) take the
form
\begin{align}
B^0 \equiv F^0_{12} = \sqrt{\frac{2}{N}}\frac{1}{r}\partial_r
f \, , \;\;\;\;\;\;
E^0 \equiv \sqrt{(F^0_{01})^2 + (F^0_{02})^2} =
\frac{1}{2N}\partial_r f_0 \ .
\end{align}
Let us study the solutions of equations (\ref{non1})-(\ref{non6}). We can
distinguish two cases: first, when the $U(1)$ and $SU(N)$ coupling
constants are equal and then the effective symmetry group is $U(N)$, and
second, when the $U(1)$ and $SU(N)$ coupling constants are different.

\subsection{$U(N)_{gauge}\times SU(N)_{global}$ solutions}

If the $U(1)$ and $SU(N)$ coupling constants are equal $\kappa_1 =
\kappa_2 \equiv \kappa$, the system of equations
(\ref{non1})-(\ref{non6}) decouples into
\begin{align}
r\partial_r\phi &= -\epsilon g \phi &
\frac{1}{r}\partial_r g &= -\frac{\epsilon}{8\kappa^2}
\phi^2(\phi^2-1)\ , \label{Trivial}\\
r\partial_r\phi_N &= -\epsilon g^{N^2-1} \phi_N &
\frac{1}{r}\partial_r g^{N^2-1} &= -\frac{\epsilon}{8\kappa^2}
\phi_N^2 (\phi_N^2 - 1) \ , \label{Nontrivial}
\end{align}
where we have defined
\begin{align}
g = \frac{1}{N}\left(f - f^{N^2-1}\right) \, , \;\;\;
g^{N^2-1} = \frac{1}{N}\left(f + (N-1) f^{N^2-1}\right) .
\label{Nontrivial2}
\end{align}
In terms of the new functions, the boundary conditions are
\begin{align}
&g(0) = 0\,,  & &g(\infty) = 0\,,  & &g^{N^2-1}(0) = -\epsilon n\,,
&  &g^{N^2-1}(\infty) = 0\label{g1}\\
&\phi(0) = C\,,  & &\phi(\infty) = 1\,, & &\phi_N(0)=0\,,
& &\phi_N(\infty) = 1 \ .\label{phi2}
\end{align}
Since the field $\phi$ has no winding, it does not necessarily
vanish at the origin so $C$ is an arbitrary parameter.
Systems (\ref{Trivial}) and (\ref{Nontrivial}) are formally the
same but the functions $(\phi, g)$ and $(\phi_N, g^{N^2 - 1})$
obey different boundary conditions. Remarkably, both systems
coincide with those arising in the abelian case discussed in
\cite{JLW}. For the pair $(\phi, g)$, boundary conditions imply
that
\begin{equation}
\phi \equiv 1 \;, \quad g \equiv 0 \ .
\end{equation}
The system (\ref{Nontrivial}) for the functions $(\phi_N, g^{N^2 - 1})$
was solved numerically in \cite{JW}. We remark that these equations do
not depend on $N$. At the origin, $\phi_N$ approaches to zero as $r^{n}$
(this fact will be important for the semi-local vortex).

%%%%%%%%%%%%%%%%%%%%%%%%%%%%%%%%%%%%%%%%%%%%%%%%%%%%%%%%%%%%%%%%%%%%%%%%%
%%%%%%%%%%%%%%%%%%%%%%%%%%%%%%%%%%%%%%%%%%%%%%%%%%%%%%%%%%%%%%%%%%%%%%%%
\begin{figure}
\centering
\includegraphics[width=16cm]{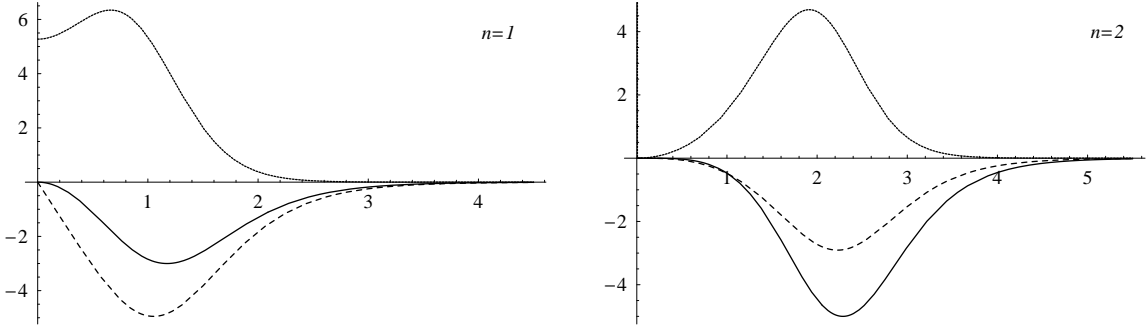}
\caption{\small Plot of the magnetic field $B^0$ (solid line),
electric field  $E^0$ (dashed line), and energy density ${\cal H}$
(dotted line) for local vortices with $\kappa_1/\kappa_2 = 1$. The
distance between the energy density maximum and the origin increases
with the winding number $n$.} \label{Bkk1}
\end{figure}
%%%%%%%%%%%%%%%%%%%%%%%%%%%%%%%%%%%%%%%%%%%%%%%%%%%%%%%%%%%%%%%%%%%%%%%%
%%%%%%%%%%%%%%%%%%%%%%%%%%%%%%%%%%%%%%%%%%%%%%%%%%%%%%%%%%%%%%%%%%%%%%%%

We show in figure \ref{Bkk1} profiles of $B^0$, $E^0$ and the energy
density ${\cal H}(r)$. As usually happens in CS theories, the sixth
order potential makes the maximum of the magnetic field to be away
from the origin.

\subsection{$ U(1)_{gauge} \times SU(N)_{gauge} \times SU(N)_{global}$
solutions}

\begin{figure}
\centering
\includegraphics[width=16cm]{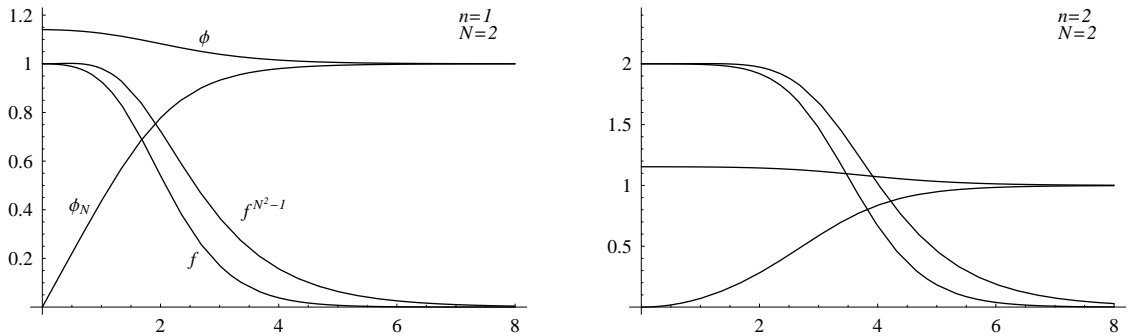}
\caption{\small Profile functions   for local vortices, for negative
magnetic flux and  $\kappa_1/\kappa_2 = 1/2$.} \label{kk05}
\end{figure}

When  $\kappa_1 \neq \kappa_2$, the complete set of equations
(\ref{non1})-(\ref{non6}) has to be solved numerically. We used a
relaxation method to find explicit numerical solutions. The ratio of
coupling constants $k = \kappa_1/\kappa_2$ is in fact the only
independent parameter of the theory, since $\kappa_1$ or $\kappa_2$ can
be absorbed by a rescaling. In fact the energy can be expressed in terms
of $k$ only. We observe that as $k$ departs from $1$, $f$ and $f^{N^2 -
1}$ tend to separate from each other as well, forcing $\phi$ to be
non-constant. As $k$ goes from $k<1$ to $k>1$, the difference
$f-f^{N^2-1}$ changes sign, forcing the derivative of $\phi$ to change
sign. Qualitatively, the behavior resembles the $U(N)$ case (in which
both coupling constant coincide). When varying the winding number $n$ of
the vortex, the profile functions change in a similar way as they do in
the $U(N)$ case. For equal $n$ and different $N$, solutions do not change
considerably. We present in figure \ref{kk05} some solutions for the case
$k =1/2$.

%%%%%%%%%%%%%%%%%%%%%%%%%%%%%%%%%%%%%%%%%%%%%%%%%%%%%%%%%%%%%%%%%%%%%%%%
%%%%%%%%%%%%%%%%%%%%%%%%%%%%%%%%%%%%%%%%%%%%%%%%%%%%%%%%%%%%%%%%%%%%%%%%
%%%%%%%%%%%%%%%%%%%%%%%%%%%%%%%%%%%%%%%%%%%%%%%%%%%%%%%%%%%%%%%%%%%%%%%%
%%%%%%%%%%%%%%%%%%%%%%%%%%%%%%%%%%%%%%%%%%%%%%%%%%%%%%%%%%%%%%%%%%%%%%%%

\section{Non-abelian semi-local strings}

We have discussed non-Abelian vortex solutions in a $U(1)_{gauge}\times
SU(N)_{gauge}\times SU(N)_{flavor}$ theory which are usually called {\it
local} vortex solutions. In order to have {\it semilocal} vortex
solutions, those for which $N_f >N$, one has to extend the matter content
of the theory by adding $N_e$ extra flavors so that $N_f = N + N_e$
\cite{AchV}-\cite{shifmansemi}. For definiteness we take $N_e$ equal to
$N$ (but a general case can be equally treated). Then, the symmetry of
the model is $U(1)_{gauge}\times SU(N)_{gauge}\times SU(2N)_{flavor}$. We
call ${\cal X}$ the extra matter fields.

In this case the trivial vacuum state is given by
\begin{align}
& \Phi^{vac} = \diag\{1, \ldots, 1\}\; , \;\;\;\; {{\cal X}}^{vac} =
\diag\{0, \ldots, 0\}\; , \;\;\;\; A_\mu^{vac} =  0\, ,\label{vacios2}
\end{align}
and it is invariant under the following transformation
\begin{equation}
\Phi^{vac} \to V^{-1} \Phi^{vac} V \; ,
  \;\;\;  \;\;\;  {\cal X}^{vac} \to V^{-1} {\cal X}^{vac}
  \tilde V = 0 \; ,
  \;\;\;  \;\;\;  A^{vac}_\mu \to V^{-1} A^{vac}_\mu V = 0\ ,
\end{equation}
where we have chosen, as in the local case, a gauge element $U = V^{-1}$
with $V$ a $N\times N$ flavor block and we have called $\tilde V$ the
$N\times N$ flavor block acting on the extra ${\cal X}$ mater fields.

We follow now  the same steps as in the local case: first we perform a
rotation of the vacuum to find  a singular vortex configuration and then
propose an ansatz for a regular configuration which we write explicitly
\begin{align}
\Phi &= \diag \{\phi(r),\dots,\phi(r),e^{-i\varphi
n\epsilon}\phi_N(r)\} \; , \;\;\;\; \mathcal{X}  =
\diag \{\chi(r), \dots, \chi(r), \chi_N(r)\}\; ,
\nonumber\\
%%%%%%%%%%%%%%%%%%%%%%%%%%%%%%%%%%%%%%%%%%%%%%%%%%%%%%%%%%%%%%%%%%%%
A_i &= \frac{i}{N}\, \varepsilon_{ij}
\frac{x_j}{r^2}(n\epsilon + f^{N^2-1}(r))
\, \diag \{1,1,\dots,1-N\} - \frac{i}{N}\,
\varepsilon_{ij}\frac{x_j}{r^2}(n\epsilon + f(r)) \,\Id \; ,
\nonumber\\
%%%%%%%%%%%%%%%%%%%%%%%%%%%%%%%%%%%%%%%%%%%%%%%%%%%%%%%%%%%%%%%%%%%%
A_0 &= \frac{i}{2N}\,
f_0^{N^2-1}(r)\, \diag\{1,1,\dots,1-N\}
+ \frac{i}{2N}\, f_0 (r)\, \Id
\label{AnsatzChi}
\end{align}
where $\Id$ is the $N\times N$ identity matrix. It should be stressed
that more general ansatz could lead to solutions which exhaust the number
$N_\rho$ of collective coordinates related to  the transverse moduli
space \cite{shifmansemi}. Our ansatz corresponds to  just two collective
coordinates.

Substituting (\ref{AnsatzChi}) into the Bogomolny equations, we arrive to
a system of six differential equations for the fields $\phi$, $\phi_N$,
$\chi$, $\chi_N$, $f$ and $f^{N^2-1}$, and two constraints for the fields
$f_0$ and $f_0^{N^2-1}$.
The solutions to these equations will have the same energy and flux than
the local ones (\ref{flux-energy}), provided they satisfy the same
boundary conditions (\ref{boundcond}). At infinity the solutions should
reach the vacuum state, so we impose
\[
\phi(\infty) = \phi_N(\infty) = 1, \ \ \ \ \chi(\infty) = \chi_N(\infty) = 0
\ ,
\]
\begin{equation}
f_0(\infty) =
f_0^{N^2-1}(\infty) = f(\infty) = f^{N^2-1}(\infty) = 0
\ .
\end{equation}

Again we distinguish the case in which the $U(1)$ and $SU(N)$
gauge coupling constants are equal from the one in which they are
different.

\subsection{$ U(N)_{gauge}\times SU(2N)_{flavor}$ semilocal
solutions}

As it happens for local vortex, when $\kappa_1 = \kappa_2 \equiv
\kappa$ the equations of motion decouple into two sets of
independent equations. One corresponds to a system with no winding
\begin{align}
&r\partial_r \phi = -\epsilon g \phi\; , \;\;\;\;\;
r\partial_r \chi = -\epsilon g \chi
\nonumber\\
&\frac{1}{r}\partial_r g = - \frac{\epsilon
}{8\kappa^2}\left(\phi^2 + \chi^2\right)\left(\phi^2 + \chi^2 -
1\right) \ , \label{semitrivial}
\end{align}
and the other corresponds to a system with winding $n$
\begin{align}
&r\partial_r \phi_N = -\epsilon g^{N^2 - 1} \phi_N\; , \;\;\;\;\;
r\partial_r \chi_N = -\epsilon\left(g^{N^2 - 1} + \epsilon n
\right)\chi_N
\nonumber\\
&\frac{1}{r}\partial_r g^{N^2 - 1} = - \frac{\epsilon
}{8\kappa^2}\left(\phi_N^2 + \chi_N^2\right)\left(\phi_N^2 +
\chi_N^2 - 1\right)  \label{seminontrivial}
\end{align}
(we have defined $g$ and $g^{N^2 - 1}$ as in (\ref{Nontrivial2})).

As in the case of local strings, these equations coincide with
those that arise in the abelian case \cite{Khare}. The system
(\ref{semitrivial}) admits a trivial solution
\begin{align}
g = \chi= 0 \; ,\;\;\;\;\;\phi =1 \ ,
\end{align}
while (\ref{seminontrivial}) can be solved numerically. Combining the
equations for $\chi_N$ and $\phi_N$, we get
\begin{equation}
\chi_N = \rho\ \frac{\phi_N}{r^n} \ ,\label{chiN}
\end{equation}
where $\rho$ is a (complex) integration constant. So, we finally have
\begin{align}
r\partial_r \phi_N &= -\epsilon g^{N^2 - 1} \phi_N
\nonumber\\
\frac{1}{r}\partial_r g^{N^2 - 1} &= - \frac{\epsilon
}{8\kappa^2}\left(\left(1 + \frac{\rho^2}{r^{2n}}\right) \phi_N^2
- 1\right)\left(1 + \frac{\rho^2}{r^{2n}}\right)\phi_N^2 \ .
\label{seminontrivial-2}
\end{align}
The same argument applied to $\chi$ and $\phi$ determines that $\chi =
\alpha \phi$, but since $\phi(\infty) = 1$ and $\chi(\infty) = 0$,
$\alpha$ must vanish, and so $\chi = 0$ everywhere. The solutions we
obtain are then the most general ones. The energy is independent of the
complex parameter $\rho$, which is then associated with the ``size
moduli''. One has has in general, $n_i$ complex parameters $\rho_i$ for
arbitrary flavor $N_f >N$ . For elementary strings one can see that $n_i
= N_f-N$ (Since we are considering $N_f = N +1$, we have in this case
just one complex parameter).

\begin{figure}
\centering
\includegraphics[width=10cm]{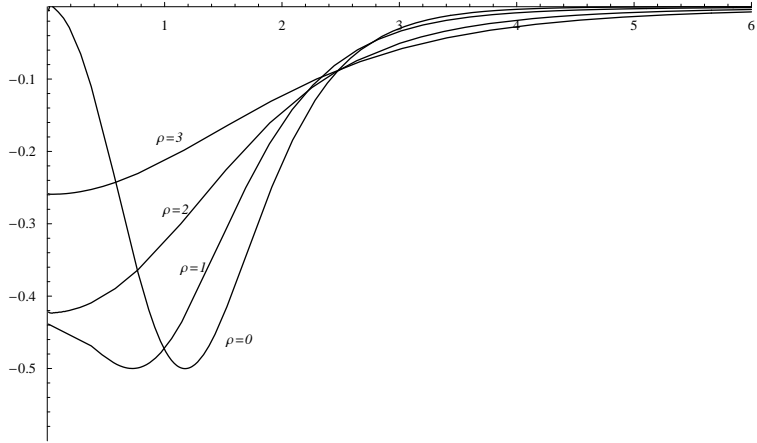}
\caption{\small  The magnetic field for semi-local vortices with
$\kappa_1 =  \kappa_2$  for different $\rho$ values .} \label{semikk1-2}
\end{figure}

From figure \ref{semikk1-2} we see that the solutions spread when the
parameter $|\rho|$ is incremented. As $\rho$ increases, since the flux is
conserved, the extremum of the magnetic field must approach to the origin
to compensate the spread towards spatial infinity. A similar phenomenon
occurs with the energy.

%%%%%%%%%%%%%%%%%%%%%%%%%%%%%%%%%%%%%%%%%%%%%%%%%%%%%%%%%%%%%%%%%%%%%%%%
%%%%%%%%%%%%%%%%%%%%%%%%%%%%%%%%%%%%%%%%%%%%%%%%%%%%%%%%%%%%%%%%%%%%%%%%
%%%%%%%%%%%%%%%%%%%%%%%%%%%%%%%%%%%%%%%%%%%%%%%%%%%%%%%%%%%%%%%%%%%%%%%%
%%%%%%%%%%%%%%%%%%%%%%%%%%%%%%%%%%%%%%%%%%%%%%%%%%%%%%%%%%%%%%%%%%%%%%%%

%\newpage

\subsection{$U(1)_{gauge} \times SU(N)_{gauge}\times SU(N+1)_{flavor}$
semilocal solutions}

We consider now the case in which  $\kappa_1 \ne  \kappa_2$. The same
arguments as above lead to $\chi(r)= 0$, so that the flavor group is in
fact reduced from $SU(2N)$ to $SU(N+1)$. Similarly, the relation
\eqref{chiN} holds, and the differential equations do not decouple. The
system is very analogous to the system (\ref{non1})-(\ref{non6}) obtained
for the local case, with the only difference that, in the equations for
the $f$'s, the field $\phi_N$ gets locally scaled
\begin{equation}
\phi_N^2 \rightarrow \phi_N^2 \left(1 + \frac{\rho^2}{r^{2n}}\right)
\ .
\end{equation}
Then, semi-local solutions arising from these equations are similar to
those shown in figure \ref{kk05} for the local case, with the only
difference that they are smoother since they decay as powers of $r$ at
spatial infinity.

~

%%%%%%%%%%%%%%%%%%%%%%%%%%%%%%%%%%%%%%%%%%%%%%%%%%%%%%%%%%%%%%%%%%%%%%%%
%%%%%%%%%%%%%%%%%%%%%%%%%%%%%%%%%%%%%%%%%%%%%%%%%%%%%%%%%%%%%%%%%%%%%%%%

In summary, our main task in this work was to solve the BPS
equations for a non-Abelian Chern-Simons-Higgs theory. By proposing
an axially symmetric ansatz we obtained non-Abelian vortex solutions
and discussed their properties. A class of vortex solutions in
non-Abelian CS theories were already known
\cite{VegaSch}-\cite{VegaSch2},\cite{Lozano}, \cite{Cugliandolo}.
The solutions discussed here correspond to more fundamental vortices
in the sense that they are genuinely non-Abelian while the former
correspond to $Z_N$ vortices with the gauge flux in the Cartan
algebra of $SU(N)$).

The model discussed here is  indeed related  to the one
analyzed by \cite{Auzzi}-\cite{GSY} except that in our case  the dynamics of
the gauge fields is governed by a CS action instead of a
Yang-Mills one. This drastically
changes the vortex properties, in particular forcing them to carry electric
charge.

In the case of local vortices, our solutions  generalize  those
discussed in  \cite{aldro} to the case in which the gauge group is
$U(1)\times SU(N)$, with the Abelian and non-Abelian sectors
having different gauge coupling constants. When both couplings are
equal, the equations decouple into two sets of equations that
coincide with those arising in the Abelian case. This is not the case when the
couplings are different and the BPS equations do not decouple.
Nevertheless we were able to construct explicit solutions and
discuss their properties.

Further, we have also considered semi-local vortices, by allowing the
flavor number $N_f$ to be larger than the color number $N_c$. As already
noted for the  Yang-Mills-Higgs system, the main feature in this case is
that the solutions develop an additional moduli $\rho$ related to the
vortex transverse size, thus modifying the asymptotic behavior of the
fields, from exponential decay for the case of local vortices, to a decay
as negative power of the radial coordinate for the semi-local ones.
Interestingly enough, one can see from our explicit solutions how the
size of the vortex grows with $\rho$.

%%%%%%%%%%%%%%%%%%%%%%%%%%%%%%%%%%%%%%%%%%%%%%%%%%%%%%%%%%%%%%%%%%%%%%%%%%%%%
%%%%%%%%%%%%%%%%%%%%%%%%%%%%%%%%%%%%%%%%%%%%%%%%%%%%%%%%%%%%%%%%%%%%%%%%%%%%%
%%%%%%%%%%%%%%%%%%%%%%%%%%%%%%%%%%%%%%%%%%%%%%%%%%%%%%%%%%%%%%%%%%%%%%%%%%%%%

\vspace{1 cm}

\noindent\underline{Acknowledgments}
We would like to thank the Sociedad Cientifica Argentina for
hospitality. We are grateful to acknowledge Le\'on Aldrovandi for
interesting discussions and comments. This work is partially
supported by CONICET (PIP6160), ANPCyT (PICT 20204), UNLP, UBA and
CICBA  grants.
%
%\newpage

%%%%%%%%%%%%%%%%%%%%%%%%%%%%%%%%%%%%%%%%%%%%%%%%%%%%%%%%%%%%%%%%%%%%%%%%%%%%%
%%%%%%%%%%%%%%%%%%%%%%%%%%%%%%%%%%%%%%%%%%%%%%%%%%%%%%%%%%%%%%%%%%%%%%%%%%%%%
%%%%%%%%%%%%%%%%%%%%%%%%%%%%%%%%%%%%%%%%%%%%%%%%%%%%%%%%%%%%%%%%%%%%%%%%%%%%%

%%%%%%%%%%%%%%%%%%%%%%%%%%%%%%%%%%%%%%%%%%%%%%%%%%%%%%%%%%%%%%%%%
\end{document}